\documentclass[fleqn,10pt]{wlscirep}

\usepackage[normalem]{ulem}
\usepackage[utf8]{inputenc}
\usepackage[T1]{fontenc}
\usepackage{physics}
\usepackage{graphicx}
\usepackage{hyperref}
\usepackage[all]{hypcap}
\usepackage[caption=false]{subfig}

\title{Spectral shaping of the biphoton state from multiplexed thermal atomic ensembles}

\author[1]{T. H. Chang}
\author[1]{G.-D. Lin}
\author[2,*]{H. H. Jen}
\affil[1]{Department of Physics, National Taiwan University, Taipei 10617, Taiwan}
\affil[2]{Institute of Physics, Academia Sinica, Taipei 11529, Taiwan}
\affil[*]{sappyjen@gmail.com}

\renewcommand{\r}{\mathbf{r}}

\renewcommand{\k}{\mathbf{k}}

\def\bea{\begin{eqnarray}}
\def\eea{\end{eqnarray}}
\def\ba{\begin{array}}
\def\ea{\end{array}}
\def\bdm{\begin{displaymath}}
\def\edm{\end{displaymath}}

\begin{abstract} 
We theoretically investigate the spectral property of a biphoton state from multiplexed thermal atomic ensembles. This biphoton state originates from the cascade emissions, which can be generated by two weak pump fields under four-wave mixing condition. Under this condition, a signal photon from the upper transition, chosen in a telecommunication bandwidth, can be generated along with a correlated idler photon from the lower infrared transition. We can spectrally shape the biphoton state by multiplexing the atomic ensembles with frequency-shifted emissions, where the entropy of entanglement can be analyzed via Schmidt decompositions. We find that this spectral entanglement increases when more vapor cells are multiplexed with correlated or anti-correlated signal and idler fields. The eigenvalues in Schmidt bases approach degenerate under this multiplexing scheme, and corresponding Schmidt numbers can be larger than the number of the multiplexed vapor cells, showing the enlarged entropy of entanglement and excess correlated modes in continuous frequency spaces. We also investigate the lowest entropy of entanglement allowed in the multiplexing scheme, which is preferential for generating a pure single photon source. This shows the potentiality to spectrally shape the biphoton source, where high-capacity spectral modes can be applied in long-distance quantum communication and multimode quantum information processing.
\end{abstract}
\begin{document}
\flushbottom
\maketitle
\section*{Introduction}

Quantum communication has been actively developed and implemented in various different platforms, which involves generation and transportation of quantum resources for the purpose of secure communication protocols \cite{Acin2018}. This secure transfer of quantum information is indispensable in realizing a broader picture of quantum internet \cite{Kimble2008} with trusted inter-connected nodes. To transmit genuine quantum entanglement to a long distance between two sites, a quantum repeater \cite{Duan2001, Sangouard2011} has been proposed and carried out as solutions to dissect the transmission distance into sections and initialize the entanglement within each section in shorter distances. This short-rang entangled states can be prepared by using quantum memories as entangled links. With local quantum measurements on the common nodes of these entangled links, the entanglement can therefore be transferred to and shared by two far sites, achieving the goal of relaying quantum entanglement. The idea of quantum repeater overcomes the harmful losses in optical fiber transmissions, and this transmission loss can be further reduced when a telecom bandwidth is applied. For typical alkali-metal atoms, the ground state transitions are in the infrared bandwidth, while the telecom transitions can be accessed via cascade atomic configurations \cite{Chaneliere2006, Radnaev2010, Jen2010}. Other atomic systems involving telecom transitions include rare-earth atoms \cite{McClelland2006, Lu2010}, erbium-doped crystal \cite{Lauritzen2010}, trapped ions \cite{Walker2018, Bock2018}, photonic crystal waveguides \cite{Eckhouse2010}, silicon nanophotonic chip \cite{Davanco2012}, nitrogen-vacancy center in diamond \cite{Dreau2018}, microring resonator \cite{Li2016}, and erbium ions in a silicon nanophotonic cavity \cite{Dibos2018}. In addition to quantum repeaters, an alternative scheme to reach global quantum communication \cite{Scheidl2013} has recently been achieved via satellite-based operations using high altitude free-space channels \cite{Yin2017, Liao2018}. 

Other than genuine and stable entanglement distribution required in transmitting information, the accessible communication capacity is another crucial element in quantum communication. Photons are the best carriers of information, whereas atoms are the best storage devices of it. With these two disparate systems, quantum correlations can be established between a spontaneously emitted photon and an atomic excitation in a Raman-type configuration. This light-matter qubit demonstrates one highly correlated entangled state in discrete spaces of light polarizations \cite{Clauser1969, Aspect1981, Kwiat1995}. Other possible continuous degrees of freedom that can be harnessed for entanglement \cite{Braunstein2005} include space \cite{Grad2012}, transverse momenta \cite{Law2004, Moreau2014}, orbital angular momenta (OAM) of light \cite{Arnaut2000, Mair2001, Molina2007, Dada2011, Agnew2011, Fickler2012, Fickler2016}, and frequencies \cite{Branning1999, Law2000, Parker2000, Jen2012-2,Jen2016a,Jen2016b,Jen2017_cascade, Li2017}. Recently, a hybrid of paths, polarizations, and OAM are exploited to constitute an 18-qubit system \cite{Wang2018_Pan}, which demonstrates a large capacity to encode quantum information. High communication capacity can thus be attainable in both discrete and continuous degrees of freedom, which enables high-dimensional control and manipulation of quantum information.

Furthermore, an enhanced performance in quantum repeater protocols has been demonstrated in multiplexing multimode quantum memories in space \cite{Collins2007, Lan2009} or time \cite{Simon2007}. This shows the advantage of multimode feature of entanglement and the capability of spectral shaping \cite{Bernhard2013, Lukens2014, Lukens2017, Kues2017}, which allows a full grip on continuous frequency correlations. With spectral entanglement in telecom bandwidths in a multiplexing scheme \cite{Jen2016a, Jen2016b}, we expect an efficient and high-capacity quantum communication and multimode quantum information processing \cite{Afzelius2009, Zheng2015}. 

\begin{figure}
\centering
\includegraphics[width=16cm,height=8.5cm]{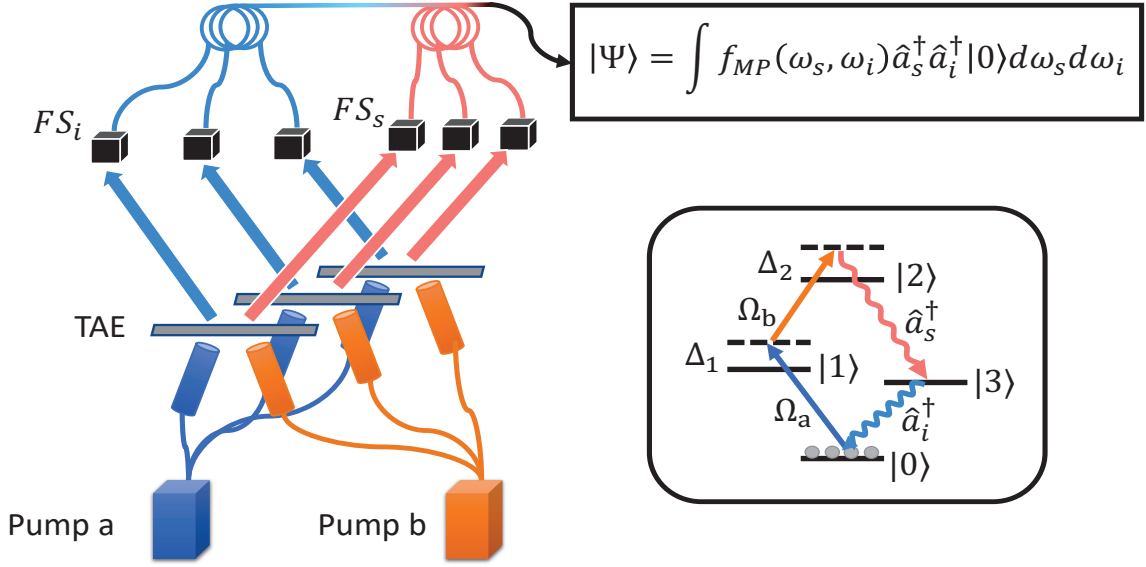}
\caption{Diagram of biphoton generation from multiplexed thermal atomic ensembles (TAEs) and diamond-type atomic configuration. The correlated biphoton source $\hat{a}^\dagger_{s,i}$ can be generated by two pump fields $\Omega_{a, b}$ with detunings $\Delta_{1,2}$. A multiplexing scheme of three TAEs are plotted for illustration, and FS$_{s(i)}$ represent frequency shifters of signal and idler photons respectively. $\ket{\Psi}$ presents the multiplexed biphoton state with an effective spectral function $f_{MP}(\omega_s,\omega_i)$.}
\label{fig1} 
\end{figure}

In this paper, we propose to use a biphoton state from multiplexed thermal atomic ensembles to realize spectral shaping of entanglement in continuous frequency spaces. As shown in Figure \ref{fig1}, multiplexing is carried out by frequency shifters, which spectrally shapes the effective biphoton spectral functions. The core element of thermal atoms in this multiplexing scheme has several advantages in quantum network \cite{Borregaard2016}. The scalability and feasibility of vapor cells \cite{Phillips2001, Balabas2010} can facilitate a quantum network with expanded inter-connected nodes. Moreover, thin vapor cells \cite{Sarkisyan2004,Keaveney2012} can even allow strong confinement of the atoms to tailor light-matter interaction strengths. Here we study the spectral property of a Doppler-broadened biphoton source in a multiplexing scheme, which is less explored in the studies of frequency entanglement. We first review the derivations of spectral functions from a Doppler-broadened biphoton source, and then formulate the effective spectral functions in a multiplexing scheme. We analyze the entropy of entanglement via Schmidt decompositions in frequency spaces, and show the increasing entanglement when the vapor cells are multiplexed along a correlation or anti-correlation directions. We finally present the lowest entropy of entanglement allowed in the multiplexing scheme, which is preferential for generating a single photon source with high purity.

\section*{Theoretical Model}\label{sec:theory}
 
As shown in Figure \ref{fig1}, highly correlated telecom and infrared photons can be spontaneously generated from two weak laser driving fields in a four-wave mixing (FWM) process. For alkali-metal atoms, the infrared transition (lower transition) is suitable for quantum storage of photons, and the telecom transition (upper transition) experiences the lowest loss in a fiber-based quantum communication. The upper excited state $|2\rangle$ can be chosen as 6S$_{1/2}$ and 4D$_{3/2(5/2)}$, or 7S$_{1/2}$ levels for rubidium or cesium atoms respectively, which lies in a bandwidth of $1.3$-$1.5$ $\mu$m \cite{Chaneliere2006}. Below we review the derivations of the coupled equations in Schr\"{o}dinger picture for a cold atomic ensemble \cite{Jen2012-2, Jen2016a, Jen2017_cascade}. Under a limit of weak excitations, we obtain the effective spectral function of a biphoton state. Then we proceed to extend the results to a thermal atomic ensemble by including Doppler broadening \cite{Chang2019_Doppler}. Following this section, we use this Doppler-broadened biphoton source as the building blocks to demonstrate the frequency entanglement manipulation and spectral shaping in a multiplexing scheme.

\subsection*{Spectral function from a cold atomic ensemble}

We use two pump fields of Rabi frequencies $\Omega_{a,b}$ respectively to drive the system from the ground state $|0\rangle$ to the upper excited state $|2\rangle$ via the intermediate state $|1 \rangle$. With FWM condition, highly correlated photons $\hat a_s$ and $\hat a_i$ are spontaneously emitted via $|3\rangle$ and back to $|0\rangle$. The Hamiltonian in interaction picture reads \cite{Jen2016a, Jen2017_cascade}
	\begin {align}
	\centering
	V_I = - \sum_{m=1,2} \Delta_m \sum_{\mu=1}^{N} \ket{m}_{\mu} \bra{m} - \sum_{m=a,b} \left(\frac{\Omega_m}{2} \hat{P}_m^\dagger + h.c.\right) - i\sum_{m=s,i} \left[ \sum_{\textbf{k}_m,\lambda_m} g_m \hat{a}_{\textbf{k}_m,\lambda_m} \hat{Q}_m^\dagger e^{-i\Delta \omega_m t} - h.c. \right], 
	\end {align}
where we let $\hbar = 1$. The detunings are defined as $\Delta_1 = \omega_a - \omega_1$ and $\Delta_2 = \omega_a + \omega_b - \omega_2$ with atomic level energies $\omega_{m = 1, 2, 3}$. The central frequencies of pump fields and emitted photons are $\omega_{a,b,s,i}$, and the polarization and wave vectors are denoted as $\lambda_m$ and $\textbf{k}_m$ respectively. We use $g_m$ as the signal and idler photon coupling constants and have absorbed $\epsilon_{\textbf{k}_m, \lambda_m} \cdot \hat{d}_m^*$ into $g_m$, where the polarization direction is $\epsilon_{\textbf{k}_m, \lambda_m}$ for quantized bosonic fields $\hat{a}_{\textbf{k}_m,\lambda_m}$ and the unit direction of dipole operators is $\hat{d}_m$. Various dipole operators are defined as $\hat{P}_a^\dagger \equiv \sum_\mu \ket{1}_\mu \bra{0} e^{i \textbf{k}_a \cdot \textbf{r}_\mu }$, $\hat{P}_b^\dagger \equiv \sum_\mu \ket{2}_\mu \bra{1} e^{i \textbf{k}_b \cdot \textbf{r}_\mu }$, $\hat{Q}_s^\dagger \equiv \sum_\mu \ket{2}_\mu \bra{3} e^{i \textbf{k}_s \cdot \textbf{r}_\mu }$, $\hat{Q}_i^\dagger \equiv \sum_\mu \ket{3}_\mu \bra{0} e^{i \textbf{k}_i \cdot \textbf{r}_\mu }$. 

When the laser fields are weak and off-resonant, satisfying $\sqrt{N} \abs{\Omega_a} / \Delta_1 \ll 1$, it is valid to assume that there is only single excitation. Therefore, we can express the state function as  
	\bea
		\ket{\psi(t)} = \mathcal{E}(t) \ket{0, vac} + \sum_{\mu = 1}^{N} A_\mu(t) \ket{1_\mu, vac} + \sum_{\mu = 1}^{N} B_\mu(t) \ket{2_\mu, vac} + \sum_{\mu = 1}^{N} \sum_{s} C_s^\mu(t) \ket{3_\mu, 1_{\textbf{k}_s, \lambda_s}} + \sum_{s,i} D_{s,i}(t) \ket{0, 1_{\textbf{k}_s, \lambda_s}, 1_{\textbf{k}_i, \lambda_i}}, 
	\eea
where the collective single excitation states and the vacuum photon state are $\ket{m_\mu} \equiv \ket{m_\mu} \ket{0}_{\nu \neq \mu}^{\otimes N-1}$ and $\ket{vac}$ respectively. By using Schr\"{o}dinger equation $i\hbar \frac{\partial}{\partial t} \ket{\psi(t)} = V_I (t) \ket{\psi(t)}$, we have the coupled equations of motion, 
	\begin{align}
		& i \dot{\mathcal{E}} = - \frac{\Omega^\ast_a}{2} \sum_\mu e^{-i \textbf{k}_a \cdot \textbf{r}_\mu} A_\mu, \\
		& i \dot{A}_\mu = - \frac{\Omega_a}{2} e^{i \textbf{k}_a \cdot \textbf{r}_\mu} \mathcal{E} - \frac{\Omega^\ast_b}{2} e^{-i \textbf{k}_b \cdot \textbf{r}_\mu} B_\mu - \Delta_1 A_\mu, \\
		& i \dot{B}_\mu = - \frac{\Omega_b}{2} e^{-i \textbf{k}_b \cdot \textbf{r}_\mu} A_\mu - \Delta_2 B_\mu - i \sum_{k_s, \lambda_s} g_s e^{i \textbf{k}_s \cdot \textbf{r}_\mu} e^{-i (\omega_s - \omega_{23} - \Delta_2) t} C^\mu_s, \\
		& \dot{C^\mu_{s,i}} = i g^\ast_s e^{-i \textbf{k}_s \cdot \textbf{r}_\mu} e^{i (\omega_s - \omega_{23} - \Delta_2) t} B_\mu - i \sum_{k_i, \lambda_i} g_i e^{i \textbf{k}_i \cdot \textbf{r}_\mu} e^{-i (\omega_i - \omega_3) t} D_{s,i}, \\
		& i \dot{D_{s,i}} = i g^\ast_i \sum_\mu e^{-i \textbf{k}_i \cdot \textbf{r}_\mu} e^{i (\omega_i - \omega_3) t} C^\mu_s,
	\end{align}
where we ignore spontaneous decays during excitation process due to large detunings. Under the adiabatic approximation, it is equivalent to solve for the steady-state solutions of these coupled equations $\dot{\mathcal{E}} = \dot{A}_\mu = \dot{B}_\mu = 0$. Since we have weak pump fields, we have $\mathcal{E} \thickapprox 1$ for the zeroth order perturbation, and $A_\mu(t)$ $\approx$ $-\Omega_a(t)e^{i\k_a\cdot\r_\mu}/(2\Delta_1)$, $B_\mu(t)$ $\approx$ $\Omega_a(t)\Omega_b(t)e^{i(\k_a+\k_b)\cdot\r_\mu}/(4\Delta_1\Delta_2)$, in the leading order of perturbations respectively.

We further consider a symmetrical single excitation state, $(\sqrt{N})^{-1}\sum_{\mu=1}^Ne^{i(\k_a+\k_b-\k_s)\cdot\r_\mu}|3\rangle_\mu|0\rangle^{\otimes N-1}$, which leads to the biphoton state $|1_{\k_s},1_{\k_i}\rangle$ generation \cite{Jen2012-2, Jen2016a, Jen2017_cascade} in a large $N$ limit,
\begin{eqnarray}
D_{s,i}(t)= g_{i}^{\ast}g_{s}^{\ast}\sum_{\mu=1}^Ne^{i\Delta\k\cdot\r_{\mu}}\int_{-\infty}^{t}\int_{-\infty}^{t^{\prime}}dt^{\prime\prime}dt^{\prime}
e^{i\Delta\omega_{i}t^{\prime}} e^{i\Delta\omega_{s}t^{\prime\prime}} \frac{\Omega_{a}(t'')\Omega_{b}(t'')}{4\Delta_{1}\Delta_{2}}e^{\left(-\Gamma_{3}^N/2+i\delta\omega_{i}\right)(t^{\prime}-t^{\prime\prime})},\label{Dsi2}
\end{eqnarray}
where $\sum_{\mu=1}^N e^{i\Delta\k\cdot\r_{\mu}}$ becomes maximal and leads to the phase-matched and highly correlated biphoton state when FWM condition $\Delta\k$ $\equiv$ $\k_{a}$ $+$ $\k_{b}$ $-$ $\k_{s}$ $-$ $\k_{i}$ $\rightarrow$ $0$ is reached. The superradiant decay constant \cite{Dicke1954, Gross1982} of the idler photon \cite{Chaneliere2006, Jen2012} is denoted as $\Gamma_{3}^{\rm N}$ $=$ $(N\bar{\mu}+1)\Gamma_{3}$ with a free-space decay rate $\Gamma_3$ and geometrical constant $\bar{\mu}$ \cite{Rehler1971}. This $\Gamma_{3}^{\rm N}$ depends on the atomic density and its geometry, which therefore can be tailored by vapor cell density and thickness. The relevant collective frequency shift \cite{Friedberg1973, Scully2009} from resonant dipole-dipole interactions \cite{Lehmberg1970} can be renormalized and absorbed into idler's central frequency. We next assume Gaussian pulse excitations $\Omega_{a,b}(t)$ $=$ $\tilde{\Omega}_{a,b}e^{-t^{2}/\tau^{2}}/(\sqrt{\pi}\tau)$ with the pulse duration $\tau$, and we obtain the probability amplitude of the biphoton state in long time limit,
	\begin {align}
	D_{si} (\Delta \omega_s, \Delta \omega_i) = \frac{\tilde{\Omega}_a \tilde{\Omega}_b g_i^* g_s^*}{4 \Delta_1 \Delta_2} \frac{\sum_\mu e^{i\Delta \textbf{k} \cdot \textbf{r}_\mu}}{\sqrt{2 \pi} \tau} f_C(\omega_s, \omega_i), ~ f_C(\omega_s, \omega_i) \equiv \frac{e^{-(\Delta \omega_s + \Delta \omega_i)^2 \tau^2 / 8}}{\frac{\Gamma_3^N}{2} - i \Delta \omega_i}, \label{eq: fc}
	\end{align}
where $\Delta \omega_s \equiv \omega_s - (\omega_2 - \omega_3 +\Delta_2)$ and $\Delta \omega_i \equiv \omega_i - \omega_3$. The joint Gaussian distribution in $f_C(\omega_s, \omega_i)$ maximizes when $\Delta \omega_s + \Delta \omega_i = 0$ which shows the energy conservation of excitations and biphoton generations, that is $\omega_s + \omega_i = \omega_a + \omega_b$.

\subsection*{Spectral function from a thermal atomic ensemble}

The spectral function from the thermal atomic ensembles can be obtained by adding Doppler broadening to the results of the cold atoms in the previous section. At some temperature $T$ of the thermal atoms, we average $f_C(\omega_s, \omega_i)$ of equation (\ref{eq: fc}) with a Maxwell-Boltzmann distribution \cite{Li1995},
	\begin {equation} \label{fd_im}
	f_D(\omega_s, \omega_i) = \int_{-\infty}^{\infty} f_C(\omega_s - k_s v, \omega_i \mp k_i v) \frac{e^{-v^2 / (2 \sigma^2)}}{\sqrt{2 \pi} \sigma} dv, 
	\end {equation}
where $\sigma \equiv \sqrt{k_B T / m}$ with $m$, the mass of the atom, and $k_B$, the Boltzmann constant. The sign of $\mp k_i v$ indicates the co-propagating or counter-propagating schemes respectively. With FWM condition, we consider only one-dimensional average of Maxwell-Boltzmann distribution. In the co-propagating scheme, we have \cite{Chang2019_Doppler}
	\begin {align} \label{fd_co}
	f_D(\omega_s, \omega_i) = \frac{-i}{\sqrt{2 \pi} \sigma k_i} e^{-\tau^2 (1-b) (\Delta \omega_s + \Delta \omega_i)^2 / 8} e^{-A^2} [\pi Erfi(A) + i \pi],~ 
	A \equiv \sqrt{\frac{\tau^2}{8b}} \ \frac{b (k_i / \bar{k}_{si}) \Delta \omega_s + (b k_i / \bar{k}_{si} - 1) \Delta \omega_i - i \Gamma^N_3 / 2}{k_i / \bar{k}_{si}}, 
	\end{align}
where $b \equiv \bar{k}^2_{si} / [\bar{k}^2_{si} + 4 / (\sigma \tau)^2]$, $\bar{k}_{si} \equiv k_s + k_i$, $k_{s, i} = |\textbf{k}_{s,i}|$, and Erfi is an imaginary error function. The effect of $\Gamma_3^N$ tends to distribute the spectral function along the anti-correlation direction (energy-conserving axis $\Delta\omega_s=-\Delta\omega_i$), and thus it increases the spectral entanglement as $\Gamma_3^N$ increases. For a smaller pulse duration $\tau$, which allows broader spectral ranges as shown in equation (\ref{fd_co}), $f_D$ has a more symmetric spectral distribution with respects to $\hat\omega_{s,i}$, leading to a less entanglement. By contrast for longer pulses, $f_D$ has more spectral weights along the correlation direction ($\Delta\omega_s=\Delta\omega_i$), which allows a larger entanglement, similar to the effect of increasing temperature \cite{Chang2019_Doppler}. These counteracting effects of preferential spectral distributions along anti-correlation or correlation directions make the spectral shaping possible to manipulate (increase or decrease) the entropy of entanglement.   

We note that our derivations here are similar to the off-resonance driving conditions in vapor cells \cite{Willis2010, Ding2012}, whereas here we further include the time-varying pump fields under weak excitation limit. This way we are able to investigate three competing energy scales on the spectral functions between excitation pulse durations, temperature of the atoms, and superradiant constant of the idler transition. Below we consider copropagating excitations of thermal atoms in a multiplexing scheme, which presents the most significant effect from these competing parameters on manipulations of spectral functions \cite{Chang2019_Doppler}, in contrast to the scheme of counter-propagating excitations.

\section*{The Multiplexed Scheme of biphoton state}

In a multiplexing scheme with frequency shifters in Figure \ref{fig1}, the effective biphoton state can be shaped and manipulated in frequency spaces. For $N_{MP}$ thermal atomic ensembles with common weak pump fields, the generated signal and idler fields can be individually frequency-shifted. We then obtain the effective spectral function from a multiplexed biphoton state 
\begin{equation} \label{eq:f_d_mp}
	f_{MP}(\omega_s, \omega_i) =  \frac{1}{\sqrt{\mathcal{N}}}\sum_{m=1}^{N_{MP}} f_{D}(\omega_s + \delta \omega_{s, m}, \omega_i + \delta \omega_{i, m}),
\end{equation}
where $\mathcal{N}$ represents the normalization of the state, and $\delta \omega_{s(i), m}$ denotes the respective frequency shifts of signal and idler photons. Note that multiphoton events more than two photons are suppressed since weak excitations are assumed. Below we first investigate two multiplexed thermal atomic ensembles, and study how the frequency entanglement can be modified depending on the directions of frequency shifts. Then we further study the trend of entanglement when more ensembles are multiplexed.

\subsection*{Multiplexed two thermal atomic ensembles}

Here we investigate the spectral entanglement from a multiplexing scheme with two thermal atomic ensembles. The entanglement in frequency spaces can be calculated via Schmidt decompositions, which we review in Methods. The effective biphoton state can be obtained by setting $N_{MP}=2$ in equation (\ref{eq:f_d_mp}). Throughout the article we use $\lambda_i=795$ nm, $\lambda_s=1.32~\mu$m, and $\Gamma_3=2\pi\times 5.8$ MHz of $D1$ transition as an example of rubidium atoms. We choose the system and excitation parameters as $\Gamma_3^N/\Gamma_3=5$ and $\Gamma_3\tau=0.25$ respectively for a moderate operating regime. We consider a range of $\pm 400 \Gamma_3$ for spectral distributions in the Schmidt decomposition, which should be broad enough for convergence of entanglement properties. The biphoton state can then be decomposed as $\ket{\Psi_{MP}}= \sum_{n} \sqrt{\lambda_n} \hat{a}^{\dagger}_{n,s} \hat{a}^{\dagger}_{n,i}$ with the eigenvalues $\lambda_n$, and signal (idler) photon operators $\hat{a}^{\dagger}_{n,s(i)}$ with the eigenmodes of $\psi_n(\omega_s)$ and $\phi_n(\omega_i)$ respectively. The entropy of entanglement can be calculated as $S = -\sum_{n=1}^{\infty} \lambda_n log_2{\lambda_n}$. 

\begin{figure}
	\centering
	\includegraphics[width=16cm,height=7.5cm]{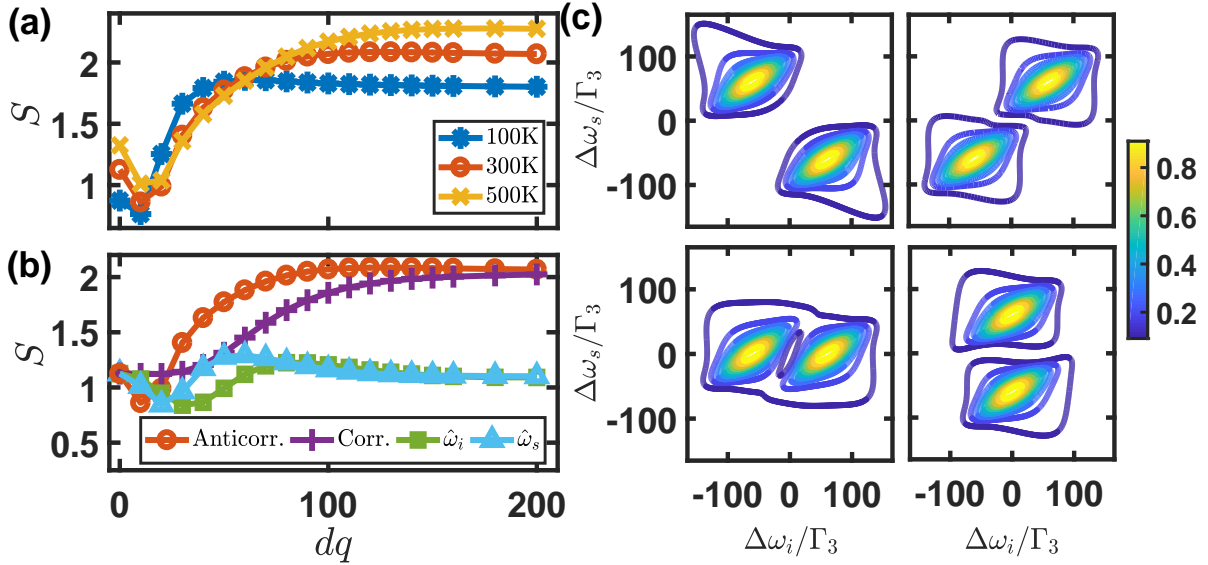}
	\caption{Entanglement and spectral distributions of multiplexed two thermal atomic ensembles (TAEs). (a) Entropy of entanglement ($S$) for multiplexed two TAEs along the direction of anti-correlation for $T = 100$K($\ast$), $300$K($\circ$), and $500$K($\cross$). The anti-correlation direction demands $\delta\omega_{s,m}=-\delta\omega_{i,m}$ with $|\delta\omega_{s,1}-\delta\omega_{s,2}|=dq$ where here $dq/\Gamma_3=120$. (b) $S$ of the biphoton state multiplexed along four different directions at $T = 300$K. We denote the correlation direction as $\delta\omega_{s,m}=\delta\omega_{i,m}$ with $|\delta\omega_{s,1}-\delta\omega_{s,2}|=dq$, $\hat\omega_i$ as $|\delta\omega_{i,1}-\delta\omega_{i,2}|=dq$ with $\delta\omega_{s,m}=0$, and $\hat\omega_s$ as $|\delta\omega_{s,1}-\delta\omega_{s,2}|=dq$ with $\delta\omega_{i,m}=0$, respectively. (c) As an example, we choose $dq/\Gamma_3=120$ and show the corresponding spectral distributions for four different directions: anti-correlation (top-left), correlation (top-right), $\hat\omega_i$ (bottom-left), and $\hat\omega_s$ (bottom-right).}\label{fig2} 
\end{figure}

In Figures~\ref{fig2}(a) and \ref{fig2}(b), we compare the entropy of entanglement $S$ at different temperatures $T$ of the atoms and for four possible multiplexing directions respectively. We define $dq$ as mutual frequency shifts between these two ensembles in respective axes of $\hat\omega_{s,i}$. At $dq=0$ in Figure \ref{fig2}(a) for various $T$, higher temperature allows larger $S$, reflecting the intrinsic entanglement property of a single vapor cell. This is expected since the spectral function broadens along the correlation direction due to Doppler effect. For larger $dq$, the asymptotic $S$ at large $dq$ increases as $T$ increases and overtakes the value at vanishing $dq$. This shows the enhanced capacity of entanglement in a multiplexing scheme. For a moderate $dq$, we see a dip first and then $S$ saturates to its asymptotic value, showing the effect of interferences between two spectral functions. This offers the capability to modify the spectral entanglement in multiplexed thermal atoms, either to enlarge or reduce it by tuning $dq$. In Figure \ref{fig2}(b), we show that both anti-correlation and correlation multiplexing schemes allow larger asymptotic $S$, which are favorable for spectral entanglement capacity, and their corresponding spectral distributions are illustrated in Figure \ref{fig2}(c). Below we further look into the effect of the number of vapor cells multiplexed in an anti-correlation direction in particular.   

\subsection*{Multiplexed multiple thermal atomic ensembles}

\begin{figure}
	\centering
	\includegraphics[width=16cm,height=7.5cm]{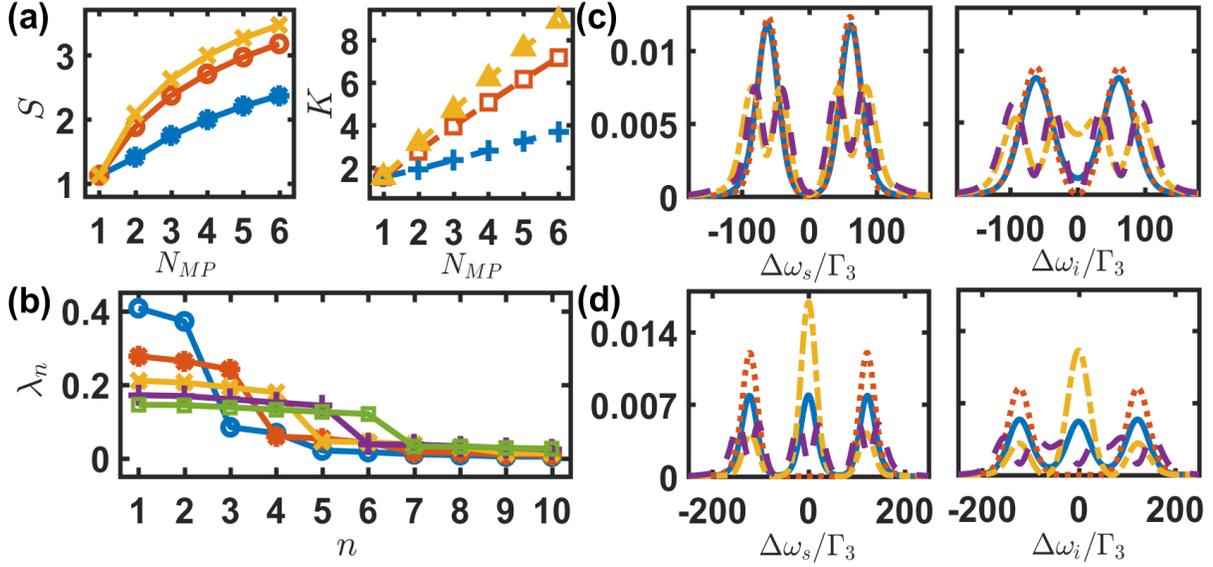}
	\caption{The entropy of entanglement $S$, Schmidt number $K$, the eigenvalues as well as the mode functions of biphoton states from multiplexed TAEs. We consider a multiplexing scheme along the anti-correlation direction and choose $T=300$K for room-temperature vapor cells. (a) We plot $S$ and $K$ with $dq/\Gamma_3 = 30$, $60$, $120$ (solid $\ast$, $\circ$, $\cross$ for $S$ and dashed $+$, $\square$, $\triangle$ for $K$). The entropy of entanglement grows logarithmically, whereas the Schmidt number grows linearly. We show the eigenvalues in (b) with $dq = 120$, and the first four mode probability densities (solid, dotted, dash-dotted, and dashed) of the signal $|\psi_n|^2$ and idler photons $|\phi_n|^2$ for two and three multiplexed TAEs in (c) and (d), respectively.}
	\label{fig3} 
\end{figure}

When more ensembles are multiplexed, we expect of a more entangled biphoton source in frequency spaces. Here we define a uniform mutual frequency shifts in a multiplexing scheme as $dq\equiv|\delta\omega_{s,m} - \delta\omega_{s,m+1}|$ between neighboring multiplexed ensembles. In Figure~\ref{fig3}(a), we show that $S$ grows logarithmically as $N_{MP}$ increases. Furthermore, we introduce a Schmidt number \cite{Law2004, Grobe1994} $K\equiv 1/\sum_n\lambda_n^2$ which represents the average correlated orthogonal modes in the system. This number presents a measure of capacity for quantum information processing, and can be directly compared to the number of multiplexed ensembles. For a small $dq/\Gamma_3\lesssim 60$, we find that $K$ is smaller than $N_{MP}$, which indicates that exploitable correlated spectral modes is less than the number of ensembles we intend to multiplex. This lack of performance is due to the interferences of these spectral functions, which can be attributed to the entanglement dip in Figure \ref{fig2}(a). On the contrary, $K>N_{MP}$ at a larger $dq$ and reaches its maximum when $dq\rightarrow\infty$ as if the ensembles are independently multiplexed. The difference of $(K-N_{MP})$ denotes an approximate excess of correlated orthogonal modes supported in the biphoton source, which is vanishing in the entanglement in discrete degrees of freedom. As a demonstration, a maximally-entangled W state in $N_{MP}$ discrete modes reads
\bea
|W\rangle=\frac{1}{\sqrt{N_{MP}}}\sum_{m=1}^{N_{MP}} \hat a_m^\dag|0\rangle,
\eea 
which leads to $K=N_{MP}$. Therefore, $K$ in Figure \ref{fig3}(a) also reflects how much the excess capacity enabled in the multiplexing scheme.

In Figure \ref{fig3}(b), we show the distribution of eigenvalues for a relatively large mutual frequency shift. Approximate degenerate eigenvalues are emerging when more number of thermal atomic ensembles are multiplexed, leading to a more entangled biphoton source. Meanwhile, the eigenmodes of signal and idler photons are shown in Figures \ref{fig3}(c) and \ref{fig3}(d), which present a feature of multiple peaks that depend on $N_{MP}$. Higher $n$ modes manifest more peaks due to the orthogonality relations, but with less weights of $\lambda_n$. These modes in frequency domains represent the main functional forms that can be used in frequency encoding or decoding in quantum key distribution \cite{Gisin2012} or to carry out Hadamard gates \cite{Lukens2014}. 

\section*{Spectral Shaping}

Finally, we explore the possibility to manipulate the spectral entanglement with a combination of four possible multiplexing directions discussed in previous sections, which is related to generation of a pure single photon source. Single photon source can be obtained by annihilating one of the biphoton source via single photon detections. This leads to a on-demand single photon generation from the conditional measurement. However, the more entangled a biphoton source is, the less pure a projected single photon becomes. Therefore, we raise the question whether it is possible to generate a less entangled biphoton source via spectral shaping in our proposed multiplexing scheme. 
     
\begin{figure}
	\centering
	\includegraphics[width=16cm,height=7.5cm]{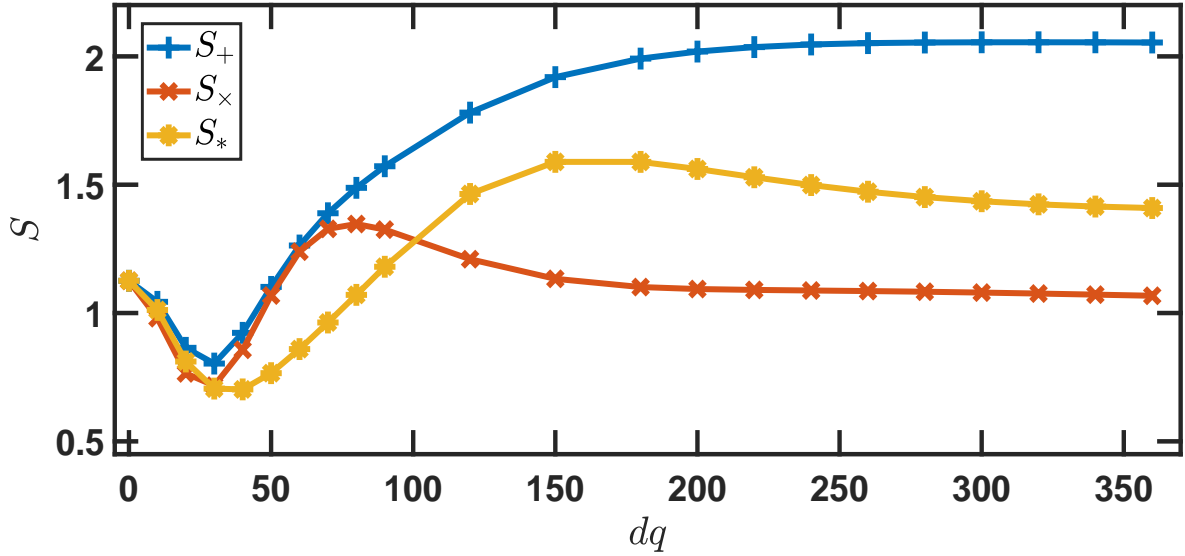}
	\caption{Spectral shaping with four and eight TAEs. We calculate spectral entanglement $S$ of three types of multiplexed spectral function as a function of $dq$. Using room-temperature vapor cells of $T = 300$K, we multiplex two pairs of TAEs, with one pair along $\hat\omega_s$ and the other $\hat\omega_i$, to get a '$+$'-type spectral function. The '$\cross$'-type can be obtained by a combination of correlation and anti-correlation multiplexing scheme. Both '$+$'- and '$\cross$'-types form a multiplexed four TAEs. To form a multiplexed eight TAEs placed on the vertices of a regular octagon, we use a combination of '$+$'- and '$\cross$'-type spectral functions and denote its spectral entanglement as $S_*$.}
	\label{fig4} 
\end{figure}

Here we integrate four possible multiplexing directions of correlation, anti-correlation, $\hat\omega_s$, and $\hat\omega_i$ to demonstrate the capability of spectrally shaping a less entangled biphoton source, preferential for creating single photon sources with high purity. For a well defined photon spectral function, its functional form should be convergent and vanishing at infinite frequency ranges. In a minimal set of parameters to describe a photon wave packet, a line width is enough to characterize the basic property of it in either a Lorentzian or Gaussian functional form, which is symmetric to its central frequency. A joint spectral function from a separable (non-entangled) biphoton state should be factorizable as $f_D(\omega_{s},\omega_{i})=g_D(\omega_{s})h_D(\omega_{i})$, and therefore the spectral function with a less $S$ should behave symmetrically with respects to the axes of $\hat\omega_s$ and $\hat\omega_i$. Based on this symmetric consideration, in Figure \ref{fig4} we use four and eight vapor cells in a multiplexing scheme as an example, and design the placements of these multiplexed vapor cells on the vertices of a square or a regular octagon in frequency axes. The mutual frequency shifts $dq$ are made equal to respective long diagonals. A scheme of multiplexed four vapor cells can be realized by superposing four spectral functions on the axes of $\hat\omega_s$ and $\hat\omega_i$, each with two multiplexed ensembles respectively. We denote it as '$+$'-type multiplexing scheme. Similarly a '$\cross$'-type spectral function can be formed by a combination of correlation and anti-correlation multiplexing schemes. To form a multiplexed eight TAEs, we use a combination of '$+$'- and '$\cross$'-type spectral functions. 

As expected, all $S$ in three types of multiplexing collapse to one another when $dq\rightarrow 0$, retrieving the result of single vapor cell. Similar dips for all these multiplexing schemes emerge as in Figure \ref{fig2} for a moderate $dq$, providing the lowest possible $S$ in our designated multiplexing schemes. As discussed in Figure \ref{fig3}, the spectral entanglement should increase when more vapor cells are multiplexed along the correlation or anti-correlation directions. By contrast, we are able to reduce the entanglement when we symmetrically multiplex these spectral functions, where $S_{+,\cross}\sim 0.75$ at $dq/\Gamma_3\sim 30$ in Figure \ref{fig4}, comparing with $S\sim 2.0$ at the same $dq$ with $N_{MP}=4$ in Figure \ref{fig3}(a). For asymptotic spectral entanglement at large $dq$, we find that $S_+$ is much larger than $S_{\cross}$, which indicates that a '$\cross$'-type multiplexing scheme is favored to reduce the entanglement of the multiplexed vapor cells. This is due to the orientation of the spectral weights aligning along the correlation direction predominantly as shown in Figure \ref{fig2}(c), which is thus can be made to be more symmetric in a '$\cross$'-type scheme. For the case of multiplexed eight thermal atomic ensembles, we find an even smaller $S_*\sim 0.7$ at $dq/\Gamma_3\sim 40$, showing a diminishing trend that favors a more pure single photon source. This demonstrates the ability of spectral shaping by multiplexing the thermal systems in symmetric orientations, which potentially can generate single photons with high purity, useful for large-scale quantum information processing \cite{Dusanowski2019}. 

\section*{Conclusion}

In conclusion, we have obtained the spectral functions of a biphoton source from thermal atomic ensembles in a frequency-multiplexing scheme. We study its spectral entanglement by Schmidt decompositions, and find that the entanglement increases when more vapor cells are multiplexed along a correlation or anti-correlation direction between signal and idler photon central frequencies. The calculated Schmidt numbers can be larger than the number of the multiplexed atomic ensembles, showing the enlarged and excess correlated modes supported in continuous frequency spaces. We also investigate the lowest entropy of entanglement allowed in the multiplexing scheme, which favors a single photon generation with high purity. We demonstrate the capability to spectrally shape the biphoton source, where high-capacity spectral modes, large scalability of vapor cells, along with low-loss telecom bandwidths, provide a superior platform to implement long-distance quantum communication and multimode quantum information processing.  

\section*{Methods}

\subsection*{Schmidt decomposition}

Here we review Schmidt decomposition in continuous frequency spaces \cite{Law2000}. We use Schmidt decomposition to analyze the spectral entanglement of the Doppler-broadened biphoton state. For a spectrally entangled biphoton generation of signal $\hat a_{\lambda_{s}}$ and idler $\hat a_{\lambda_{i}}$ photons with some polarizations $\lambda_{s}$ and $\lambda_{i}$ respectively, we can express the biphoton state $|\bar\Psi\rangle$ with a spectral function $\bar f(\omega_{s},\omega_{i})$,
\begin{equation}
|\bar\Psi\rangle=\int \bar f(\omega_{s},\omega_{i})\hat{a}_{\lambda_{s}}^{\dag}(\omega_{s})\hat{a}_{\lambda_{i}}^{\dag}(\omega_{i})|0\rangle d\omega
_{s}d\omega_{i}.
\end{equation}
We can quantify the entropy of entanglement of the spectrally entangled biphoton state in the Schmidt bases, where the state vectors can be written as
\begin{eqnarray}
|\bar\Psi\rangle&=&\sum_{n}\sqrt{\lambda_{n}}\hat{b}_{n}^{\dag}\hat{c}_{n}^{\dag}|0\rangle,\\
\hat{b}_{n}^{\dag}&\equiv&\int\psi_{n}(\omega_{s})\hat{a}_{\lambda_{s}}^{\dag}(\omega_{s})d\omega_{s},\\
\hat{c}_{n}^{\dag}&\equiv&\int\phi_{n}(\omega_{i})\hat{a}_{\lambda_{i}}^{\dag}(\omega_{i})d\omega_{i}.
\end{eqnarray}
The effective creation operators $\hat{b}_{n}^{\dag}$ and $\hat{c}_{n}^{\dag}$ associate with the eigenmodes $\psi_{n}$ and $\phi_{n}$ respectively, and $\lambda_n$'s are the eigenvalues and probabilities for the $n$th eigenmode. These eigenmodes can be obtained by
\begin{eqnarray}
&&\int K_{1}(\omega,\omega^{\prime})\psi_{n}(\omega^{\prime})d\omega^{\prime}  =\lambda_{n}\psi_{n}(\omega),\\
&&\int K_{2}(\omega,\omega^{\prime})\phi_{n}(\omega^{\prime})d\omega^{\prime}  =\lambda_{n}\phi_{n}(\omega),
\end{eqnarray}
where the kernels for one-photon spectral correlations \cite{Law2000, Parker2000} can be constructed as
\begin{eqnarray}
&&K_{1}(\omega,\omega^{\prime}) \equiv\int f_D(\omega,\omega_{1})f_D^{\ast}(\omega^{\prime},\omega_{1})d\omega_{1},\\
&&K_{2}(\omega,\omega^{\prime}) \equiv\int f_D(\omega_{2},\omega)f_D^{\ast}(\omega_{2},\omega^{\prime})d\omega_{2}. 
\end{eqnarray}
The orthogonality of these eigenmodes can be guaranteed since $\int\psi_{i}(\omega)$$\psi_{j}^*(\omega)d\omega$ $=$ $\delta_{ij}$ and $\int\phi_{i}(\omega)$$\phi_{j}^*(\omega)d\omega$ $=$ $\delta_{ij}$, and the normalization of Schmidt analysis thus requires $\sum_{n}\lambda_{n}$ $=$ $1$.

The Von Neumann entropy of entanglement $S$ in the Schmidt bases is then ready to be calculated as
\begin{equation}
S=-\sum_{n=1}^{\infty}\lambda_{n}\textrm{log}_2\lambda_{n},\label{entropy}
\end{equation}
where a non-entangled state with $\lambda_{1}=1$ makes a vanishing $S$, and a finite bipartite entanglement of $S>0$ shows up when more than one Schmidt numbers $\lambda_n$ are present in $|\Psi\rangle$.


\section*{Acknowledgments}

This work is supported by the Ministry of Science and Technology (MOST), Taiwan, under the Grant No. MOST-106-2112-M-001-005-MY3 and 107-2811-M-001-1524. GDL thanks the support from MOST of Taiwan under Grant No. 105-2112-M-002-015-MY3 and National Taiwan University under Grant No. NTU-106R891708. We are also grateful for the support of NCTS ECP1 (Experimental Collaboration Program).


\section*{Author contributions statement}

T. H. Chang conducted the derivations and numerical simulations; all of the authors analyzed the results and wrote the manuscript.

\section*{Additional information}

\textbf{Competing financial interests:} The authors declare that they have no competing interests. 

\noindent\textbf{Publisher's note:} Springer Nature remains neutral with regard to jurisdictional claims in published maps and institutional affiliations.

\end{document}